\documentclass[conference]{IEEEtran}
\usepackage{cite}
\usepackage{amsmath,amssymb,amsfonts}
\usepackage{algorithmic}
\usepackage{graphicx}
\usepackage{textcomp}
\usepackage{xcolor}
\usepackage{multirow}
\usepackage{url}
\usepackage{subfigure}
\usepackage{float}
\usepackage{booktabs}
\def\BibTeX{{\rm B\kern-.05em{\sc i\kern-.025em b}\kern-.08em
    T\kern-.1667em\lower.7ex\hbox{E}\kern-.125emX}}

\begin{document}

\title{Surgeons Are Indian Males and Speech Therapists Are White Females: Auditing Biases in Vision-Language Models for Healthcare Professionals\\
}

\author{\IEEEauthorblockN{Zohaib Hasan Siddiqui}
\IEEEauthorblockA{\textit{Computer Science and Engineering} \\
\textit{Jamia Hamdard}\\
New Delhi, India \\
zohaibhasan066@gmail.com}
\and
\IEEEauthorblockN{Dayam Nadeem}
\IEEEauthorblockA{\textit{Computer Science and Engineering} \\
\textit{Jamia Hamdard}\\
New Delhi, India \\
dayam8696@gmail.com}
\and
\IEEEauthorblockN{Mohammad Masudur Rahman}
\IEEEauthorblockA{\textit{Computing and Informatics} \\
\textit{University of Louisiana at Lafayette}\\
Lafayette, LA, United States \\
mohammad.rahman3@louisiana.edu}
\and
\IEEEauthorblockN{Mohammad Nadeem}
\IEEEauthorblockA{\textit{Computer Science} \\
\textit{Aligarh Muslim University}\\
Aligarh, India \\
mnadeem.cs@amu.ac.in}
\and
\IEEEauthorblockN{Shahab Saquib Sohail}
\IEEEauthorblockA{\textit{Computer Science and Engineering} \\
\textit{Jamia Hamdard}\\
New Delhi, India \\
shahabsaquibsohail@gmail.com}
\and
\IEEEauthorblockN{Beenish Moalla Chaudhry}
\IEEEauthorblockA{\textit{Computing and Informatics} \\
\textit{University of Louisiana at Lafayette}\\
Lafayette, LA, United States \\
beenish.chaudhry@louisiana.edu}
}

\maketitle

\begin{abstract}
Vision language models (VLMs), such as CLIP and OpenCLIP, can encode and reflect stereotypical associations between medical professions and demographic attributes learned from web-scale data. We present an evaluation protocol for healthcare settings that quantifies associated biases and assesses their operational risk. Our methodology (i) defines a taxonomy spanning clinicians and allied healthcare roles (e.g., surgeon, cardiologist, dentist, nurse, pharmacist, technician), (ii) curates a profession-aware prompt suite to probe model behavior, and (iii) benchmarks demographic skew against a balanced face corpus. Empirically, we observe consistent demographic biases across multiple roles and vision models. Our work highlights the importance of bias identification in critical domains such as healthcare as AI-enabled hiring and workforce analytics can have downstream implications for equity, compliance, and patient trust.\footnote{The code can be found at \url{https://github.com/zohaibhasan066/Healthcare-vlm-fairness-benchmarks}\\
\\This paper has been accepted for publication in the proceedings of 25th IEEE International Conference on Data Mining (ICDM 2025) }.
\end{abstract}


\section{Introduction}
Vision language models (VLMs) constitute a class of AI architectures that learn joint representation by aligning visual perception with natural language semantics ~\cite{nadeem2024vision}. Typically, an image encoder is paired with a text encoder and trained to inhabit a shared embedding space that supports cross-modal correspondence between images and linguistic descriptions. One such instance is OpenAI’s CLIP (Contrastive Language Image Pretraining) which is optimized on roughly 400 million image–text pairs and exhibits strong zero-shot ability for image recognition ~\cite{radford2021learning}. VLMs enable a broad spectrum of multimodal functionalities, including image captioning, visual question answering, and bidirectional text–image retrieval with downstream applications in search, recommendation, and human–computer interaction.
\begin{figure}[t]
\centering
{\includegraphics[width=0.48\textwidth]{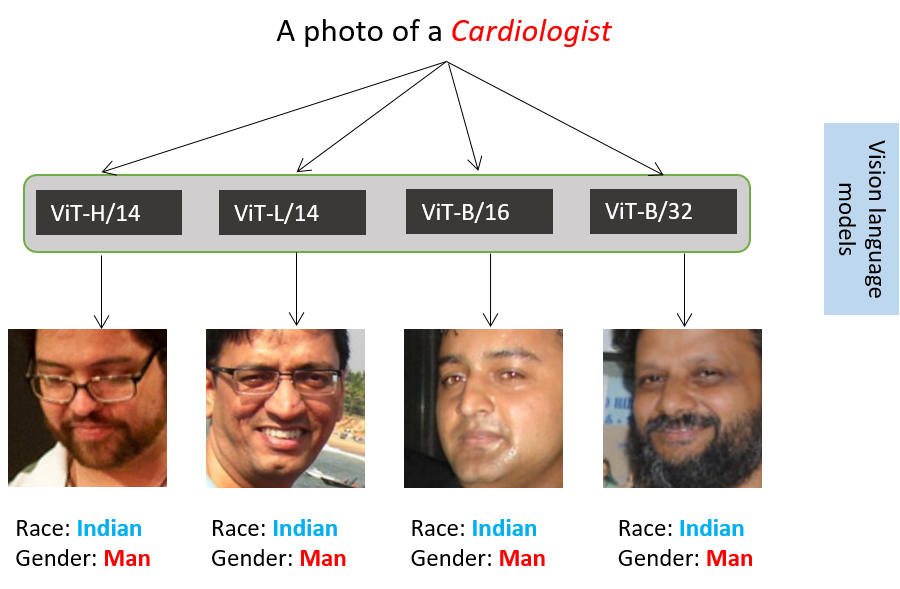}}
\caption{Gender and race bias in vision language model}
\label{Fig:Introduction}
\end{figure}
\par Although VLMs have shown remarkable abilities for various tasks, they are prone to absorbing and reproducing biases embedded in their training corpora. Empirical studies show that CLIP-like models internalize historical and societal stereotypes that appear as skewed associations in downstream outputs ~\cite{aman2025owls}. For instance, prompting with the term “terrorist” disproportionately retrieves images of Middle Eastern men ~\cite{hamidieh2024identifying}. Such errors are structured and align with popular misconceptions \cite{vo2025vision}. Researchers have documented biases in VLMs across multiple dimensions, including gender, race, age and religion. \cite{hwang2024debiasing,nadgen}. Although mitigation remains an active area of research, debiasing is challenging, as naïve interventions can degrade cross-modal alignment or reduce overall accuracy \cite{zhang2025jointCVPR}. The persistence of bias underscores the need to precisely characterize problematic behaviors and to devise targeted, context-aware interventions.

Fairness and bias in AI models for healthcare remains an active area of research. FairCLIP \cite{luo2024fairclip} conducted one of the first fairness audits of medical VLMs, documenting significant subgroup disparities in CLIP/BLIP-2 models and proposing methods to reduce non-trivial demographic bias. FairMedFM \cite{jin2024fairmedfm} and MEDFAIR \cite{zong2022medfair} systematized fairness benchmarking across medical datasets and models, revealing persistent disparities. Survey studies on CLIP in medical imaging consolidate evidence on capabilities and limitations, including fairness concerns relevant to clinical adoption \cite{zhao2025clip}. AI models are known to learn spurious correlations, leading to systematic subgroup errors and inflated performance, which reinforces the need for bias-aware evaluation and monitoring (see Fig. \ref{Fig:Introduction}). Moreover, these models are increasingly repurposed beyond diagnostics, for example, in recruitment, workforce analytics, and education, raising the risk that biases may propagate into AI-enabled hiring and representation of healthcare roles.

Despite the momentum, we find no prior study that conducts a profession-wise audit of VLMs to examine whether healthcare roles are systematically associated with particular demographics. Existing medical VLM audits focus primarily on datasets, model performance, or aggregate subgroup metrics rather than role-level semantic associations, which can adversely impact the hiring process or workforce analytics. To address this gap, our study performs a profession-sensitive evaluation of VLMs for healthcare settings. Specifically, we constructed a structured taxonomy of clinicians and allied health professionals and designed a professional-aware prompt suite to probe the behavior of VLMs. The main categories of roles were Physicians / Specialists, Nursing and Support, Technical and Laboratory, Emergency and Field, and Hospital Administration. Using the CLIP and OpenCLIP model families, we evaluated demographic bias against the FairFace benchmark dataset. Bias evaluation was performed using top-k retrieval and JS Divergence-based bias scores. As the final step, we translate the empirical findings into actionable inferences. 

\section{Related work}
Since the inception of VLMs, researchers have explored diverse applications, including image captioning, cross-modal retrieval, and multimodal reasoning, revealing not only the broad capabilities but also the persistence of systematic biases across demographic lines. Gender biases in VLMs, especially CLIP-style architectures, have manifested in occupational associations, with models disproportionately aligning high-status roles (e.g., doctor, CEO) with white men, while relegating subordinate stereotypes (e.g., nurse, caregiver) to women or marginalized groups~\cite{konavoor2025vision}. %Beyond occupational skew, studies such as ~\cite{gdelt2021clipbias} provide quantitative frameworks that embedding-based association scores reveal the process of gendered stereotype embeddings when mapping occupations and tasks onto face images.
Racial and intersectional biases are similarly salient. For instance, the authors in~\cite{baherwani2024stereotypes} have shown that CLIP’s embeddings reflect a strong influence of race and gender when interpreting demeanor and intelligence, using the FairFace dataset. Expanding the lens of modality, Hazirbas et al.~\cite{hazirbas2024bias} found that darker-skinned individuals are 4–7 times more likely to be associated with harmful labels (e.g., “cockroach,” “pig”) in CLIP and BLIP-2 models, highlighting an alarming disparity in representation.

Some of the studies have reported semantic biases. For example, Hamidieh et al.~\cite{hamidieh2024identifying} introduced the So‑B‑IT taxonomy to systematically analyze how demographic groups are associated with negative concepts, such as terrorism or criminality. Their findings showed that CLIP frequently links the concept of “terrorist” with images of Middle Eastern males, revealing how the model embeds harmful and persistent stereotypes into its representations. A similar study by Janghorbani et al.~\cite{janghorbani2023mmbias} extends the prior work on bias by curating image-text pairs across religion, nationality, sexual orientation, and disability, revealing meaningful bias across multiple subgroups in self‑supervised VLMs like ALBEF and ViLT.

Mitigation strategies have progressed alongside detection efforts, and addressing religious bias has become particularly critical. Abid et al. \cite{abid2021large} highlighted that large language and vision-language models often reproduce Islamophobic stereotypes, with prompts related to Muslims generating disproportionately negative or violent associations. This finding is supported by subsequent multimodal studies, including Raj et al.~\cite{raj2024biasdora}, which expanded the analysis to broader range of subgroups, including religion, nationality, and disability. 

Furthermore, Zhang et al applied~\cite{zhang2025jointCVPR} post-hoc adjustments to reduce gender, race, and age biases through prompt engineering and model correction, achieving measurable reductions in skewed associations. Trusted debiasing methods such as DAUDoS and counterfactual data augmentation designed to reduce stereotypicality with minimal data requirements to curb stereotypicality with minimal data requirements were also introduced in Hruday et al.~\cite{hruday2025freeze}. 

Other researchers have proposed synthetic benchmarks that span professions, age groups, racial categories, and gender identities to systematically audit bias across various VLM inference pipelines~\cite{sathe2024unified}. Intersectional and cross-modal intersectional biases have also been explored. For example, Howard et al. ~\cite{howard2023probing} demonstrated that synthetic counterfactual training can significantly reduce skew in intersectional race–gender biases without overlapping training and test subjects. 

Recent research by Vo et al. unveiled~\cite{vo2025vision} that expectations grounded in familiar semantics (e.g., brand recognition in logos) can lead to dramatic failures in objective tasks, such  as counting and illustrating how learned priors can distort the reasoning capabilities of vision-language models. Additionally, Lee et al.~\cite{lee2025visual} has shown that VLMs tend to produce more homogeneous narratives for prototypical women and White individuals, compared to more varied representations for Black Americans, revealing the differential ability to tell stories about individuals. 

Beyond these human-centric attributes, recent studies have extended the scope of bias in CLIP to non-human symbolic representations. Aman et al.~\cite{aman2025owls} has demonstrated that VLMs frequently propagate culturally ingrained animal stereotypes, such as associating owls with wisdom or foxes with unfaithfulness, even when prompts are neutral. In parallel, Anas et al.~\cite{anashorses} proposed an Animal Bias Taxonomy (ABT) to categorize cultural stereotypes about animals (e.g., donkeys as 'dumb', horses as 'strong') and use it to systematically audit CLIP against a curated dataset. 

Collectively, these studies present a robust mapping of biases in VLMs across gender, race, religion, and intersectional dimensions. They also document a range of mitigation approaches ranging from debiased prompting to counterfactual augementation to synthetic benchmarks and modality-aware interventions. However, the existing literature largely focuses on general semantic or demographic biases. Most fairness work in the medical domain (e.g., FairCLIP) centers on disease classification or clinical image analysis, rather than on role-based semantic associations specific to healthcare professions~\cite{luo2024fairclip}. Furthermore, even the profession‑centric VLM bias frameworks rarely explore healthcare-specific roles such as surgeon, nurse, technician, or pharmacist in detail. 

Our primary contribution is to fill this gap by introducing a profession-aware taxonomy of healthcare roles and benchmarking skew using a balanced face dataset (FairFace). This approach surfaces implicit societal biases in VLMs regarding who looks like a healthcare professional, which is a nuanced issue with direct implications for AI use in hiring, workforce analytics, representation, and trust within sensitive domains. 

\section{Methodology}
\begin{figure*}[!t]
    \centering
    \includegraphics[width=1\linewidth]{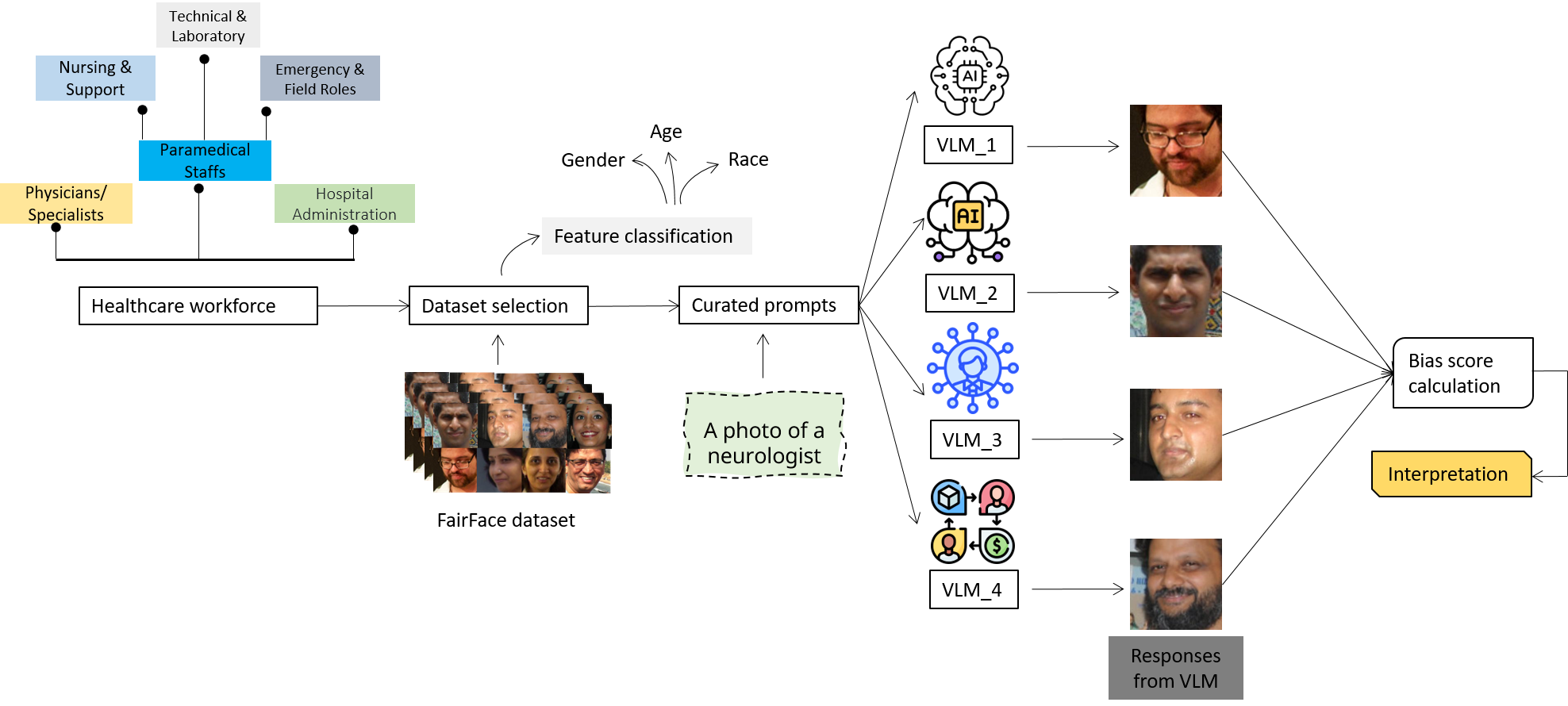}
    \caption{Overall pipeline of our adopted methodology}
    \label{fig:method}
\end{figure*}

The adopted methodology for the current work is described in this section and shown in Fig. \ref{fig:method}.

\subsection{Taxonomy}
To design the taxonomy, we considered 33 roles of the healthcare workforce and divided them into three broad categories (see Table \ref{tab:roles}). At the top are Physicians/Specialists (14 roles) including Family Doctor, Surgeon, Dentist, Orthopedic Surgeon, Cardiologist, etc. to capture the decision-making and diagnostically intensive practice areas with historical gender and age related stereotypes. Paramedical Staff are divided into Nursing \& Support (Nurse, Midwife, Nursing Assistant), Technical \& Laboratory (Pharmacist, Chemist, Laboratory Technician, Radiology Technician, Physiotherapist, etc.) and Emergency \& Field Roles (Paramedic, Ambulance Driver, Emergency Medical Technician). The paramedic category reflects patient proximity, rapid-response, and skill profiles that often anchor staffing pipelines. Lastly, Hospital Administration (Hospital Receptionist, Hospital Guard, Ward Attendant, etc.) aggregates the non-clinical but highly visible roles that shape first contact and inpatient experience. The taxonomy design allows for consistent prompting, clean stratification, and comparative analyses across clinical, allied, emergency, and administrative functions.

\begin{table*}[!t]
\centering
\small % Reduce font size
\setlength{\tabcolsep}{3pt} % Reduce column padding
\renewcommand{\arraystretch}{1.2} % Adjust row height
\caption{The proposed taxonomy of healthcare professional roles}
\label{tab:roles}
\begin{tabular}{l|l|l}
\hline
\textbf{Category} & \textbf{Sub-category} & \textbf{Professional roles} \\ \hline
Physicians/Specialists & -- & \begin{tabular}[c]{@{}l@{}}General Practitioner/Family Doctor, Surgeon, Dentist, Orthopedic Surgeon, \\ Cardiologist, Neurologist, Oncologist, Pediatrician, Gynecologist/Obstetrician, \\ Dermatologist, Radiologist, Psychiatrist, Anesthesiologist, Pathologist\end{tabular} \\ \hline
\multirow{3}{*}{Paramedical Staffs} & Nursing \& Support & Nurse, Midwife, Nursing Assistant \\ 
\cline{2-3} & Technical \& Laboratory  & \begin{tabular}[c]{@{}l@{}}Pharmacist, Chemist, Laboratory Technician, Radiology Technician, \\ Physiotherapist, Occupational Therapist, Speech Therapist\end{tabular} \\ 
\cline{2-3} & Emergency \& Field Roles & Paramedic, Ambulance Driver, Emergency Medical Technician \\ \hline
Hospital Administration & --                       & \begin{tabular}[c]{@{}l@{}}Hospital Receptionist, Hospital Guard/Security Staff, Ward Attendant, \\ Hospital Cleaner, Cafeteria Worker, Medical Records Clerk\end{tabular}                                                                                 \\ \hline
\end{tabular}
\end{table*}

\subsection{Dataset}
\begin{figure}
\centering
\subfigure[]{\includegraphics[height=0.15\textwidth, width=0.15\textwidth]{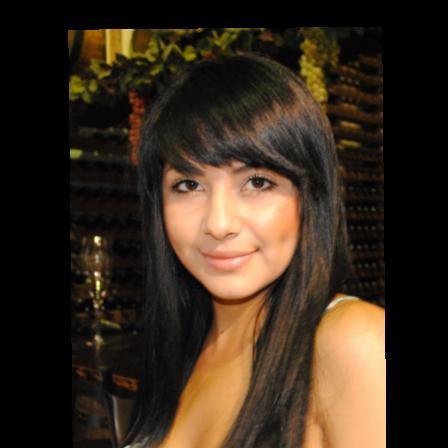}}
\hfill
\subfigure[]{\includegraphics[height=0.15\textwidth, width=0.15\textwidth]{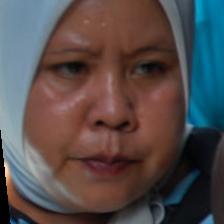}}
\hfill
\subfigure[]{\includegraphics[height=0.15\textwidth, width=0.15\textwidth]{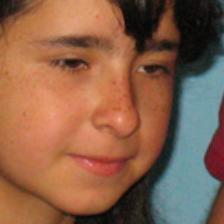}}
\\
\subfigure[]{\includegraphics[height=0.15\textwidth, width=0.15\textwidth]{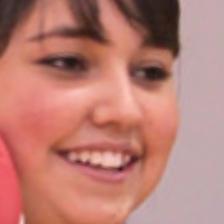}}
\hfill
\subfigure[]{\includegraphics[height=0.15\textwidth, width=0.15\textwidth]{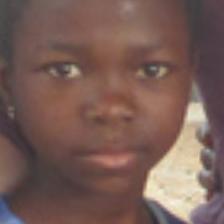}}
\hfill
\subfigure[]{\includegraphics[height=0.15\textwidth, width=0.15\textwidth]{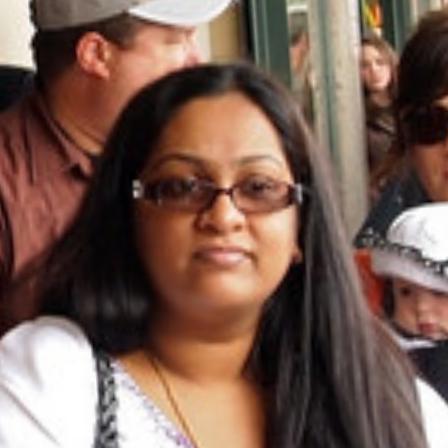}}
\\
\subfigure[]{\includegraphics[height=0.15\textwidth, width=0.15\textwidth]{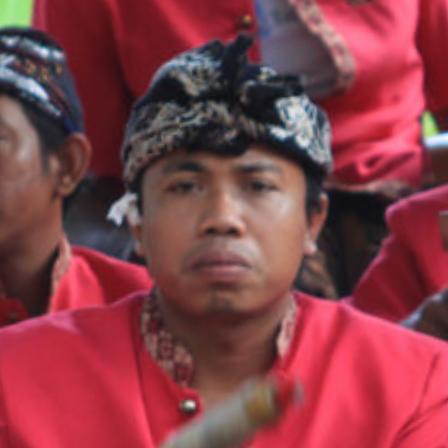}}
\hfill
\subfigure[]{\includegraphics[height=0.15\textwidth, width=0.15\textwidth]{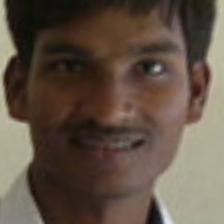}}
\hfill
\subfigure[]{\includegraphics[height=0.15\textwidth, width=0.15\textwidth]{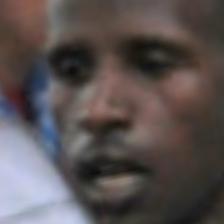}}
\\
\subfigure[]{\includegraphics[height=0.15\textwidth, width=0.15\textwidth]{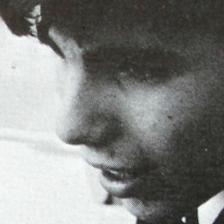}}
\hfill
\subfigure[]{\includegraphics[height=0.15\textwidth, width=0.15\textwidth]{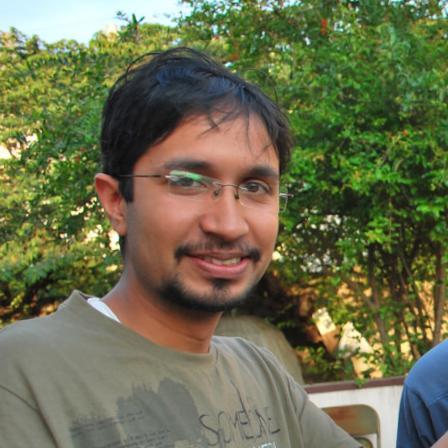}}
\hfill
\subfigure[]{\includegraphics[height=0.15\textwidth, width=0.15\textwidth]{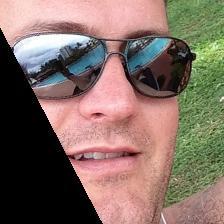}}
\caption{Representative samples from the FairFace dataset across varying demographic attributes including gender, race and age.}
\label{fig:sample}
\end{figure}

FairFace \cite{karkkainenfairface} is a large and in-the-wild face attribute dataset explicitly constructed for balanced demographic coverage. It is comprised of 108,501 images annotated for race (7 groups: White, Black, Indian, East Asian, Southeast Asian, Middle Eastern, Latino), gender, and age. The dataset was primarily sourced from YFCC-100M with additional images from media/Twitter and labeled via multi-annotator workflows. The authors of dataset showed that models trained or evaluated with FairFace achieve more consistent accuracy across demographic groups than prior datasets such as UTKFace \cite{zhang2017age} or CelebA\cite{liu2015deep}. Therefore, we used FairFace dataset due to its balanced nature. The granular race taxonomy (including Middle Eastern and distinguishing East vs. Southeast Asian) of the dataset was especially valuable for our healthcare analysis. In addition to racial diversity, dataset provides annotations for gender (male and female) and age, the latter categorized into nine groups. For our analysis, we consolidate the age dimension into three broader categories: young (0-19 years), adult (20-49 years), and old (50 years and older). A sample of the dataset is shown in Fig. \ref{fig:sample}.

\subsection{VLMs used}
We evaluated OpenAI's CLIP (ViT-B/16, ViT-B/32) and OpenCLIP (ViT-L/14, ViT-H/14) vision models for bias identification. All four models pair a Vision Transformer (ViT) image encoder with a Transformer text encoder trained to align modalities in a shared embedding space through contrastive learning \cite{cherti2023}. CLIP models are trained on 400M image-text pairs using a dual-encoder contrastive objective~\cite{radford2021learning}. While OpenCLIP models are larger and trained on LAION-2B, an open-source image-text corpus~\cite{schuhmann2022laion} Though CLIP and OpenCLIP models are similar in architecture, they are trained on different datasets which allows us to test whether measured biases are VLM family-specific or consistent phenomena across implementations. Moreover, we used off-the-shelf pretrained weights without task-specific fine-tuning to identify inherent model associations learned from large-scale web data.

\subsection{Prompt Design}
Our prompt design considers each role in the taxonomy as an atomic term to elicit model associations without explicitly mentioning any gender, age, race, attire, or equipment-related tokens. The prompt template used was: \texttt{Photo of a \{professional\}}, such as
\texttt{Photo of a dentist} or \texttt{Photo of a hospital receptionist}.  It ensures that any observed demographic bias reflects intrinsic role association priors of the VLMs rather than artifacts of phrasing or explicit attribute mentions. The occupations are derived from the taxonomy described earlier in Table \ref{tab:roles}. 

\subsection{Bias identification}
As a first step, we computed similarity using cosine distance for a given image and prompt. Further, top-$k$ retrieval analysis was utilized to quantify the bias~\cite{hamidieh2024identifying}. Given a role prompt $o$ and its text $\tau(o)$, let $f(\cdot)$ and $g(\cdot)$ denote the image and text encoders of the VLM, respectively. For any image $x$ from the evaluation set $\mathcal{X}=\{x_1,\dots,x_N\}$, we compute the cosine similarity using equation \ref{eq:cosine}.
\begin{align}
s(x,o) \;=\; \frac{\langle f(x),\, g(\tau(o))\rangle}{\|f(x)\|\,\|g(\tau(o))\|}.
\label{eq:cosine}
\end{align}

Once similarity score is calculated using equation \ref{eq:cosine}, the top-$k$ retrieved set for prompt $o$ can given defined using equation \ref{eq:topk}.
\begin{align}
\mathcal{R}_k(o) \;=\; \operatorname*{arg\,top\text{-}k}_{x\in\mathcal{X}} \, s(x,o).
\label{eq:topk}
\end{align}

For our analysis, we set $k=100$. If retrieved images are disproportionately drawn from stereotyped classes (e.g., \textit{``nurses are female''}), we consider the model to exhibit bias for that prompt. On the other hand, if the distribution of face images is uniform across the top-k retrieved images, we infer that the VLM does not exhibit bias for the specific role. This process was repeated for all prompts, and the results were analyzed to identify recurring patterns.

\subsection{Performance measures}
Apart from top-k retrieval analysis, we also derived a bias score using Jensen–Shannon (JS) divergence \cite{sutter2020multimodal}. For each professional role and model, we represented the observed demographic distribution as a probability vector as shown in equation \ref{eq:prob}.
\begin{align}
\hat{\mathbf{p}} = (p_1, p_2, \dots, p_C),
\label{eq:prob}
\end{align}
where $C$ is the number of categories (e.g., $C=2$ for gender, $C=3$ for age, $C=7$ for race), and each $p_i$ corresponds to the proportion of retrieved images belonging to category $i$, normalized such that $\sum_{i=1}^C p_i = 1$.

We defined an equal baseline distribution according to equation \ref{eq:base}.
\begin{align}
\mathbf{b} = (b_1, b_2, \dots, b_C), \quad \text{with } b_i = \frac{1}{C}, \ \forall i,
\label{eq:base}
\end{align}
which represents the ideal case of perfect demographic balance (e.g., 50/50 for gender, 33.3/33.3/33.3 for age, and $1/7$ each for race). To quantify the deviation of $\hat{\mathbf{p}}$ from $\mathbf{b}$, we employed the Jensen--Shannon (JS) divergence as given in equations \ref{eq:JS}, \ref{eq:M} and \ref{eq:D}.
\begin{align}
\text{Bias}(\hat{\mathbf{p}}, \mathbf{b}) = \frac{1}{2} D_{\mathrm{KL}}(\hat{\mathbf{p}} \,\|\, \mathbf{m}) + \frac{1}{2} D_{\mathrm{KL}}(\mathbf{b} \,\|\, \mathbf{m}),
\label{eq:JS}
\end{align}
where
\begin{align}
\mathbf{m} = \frac{1}{2}(\hat{\mathbf{p}} + \mathbf{b}),
\label{eq:M}
\end{align}
and
\begin{align}
D_{\mathrm{KL}}(p \,\|\, q) = \sum_{i=1}^C p_i \log \frac{p_i}{q_i}
\label{eq:D}
\end{align}
denotes the Kullback--Leibler divergence.

The JS divergence is symmetric, finite, and bounded (see equation \ref{eq:bound}).
\begin{align}
0 \leq \text{Bias}(\hat{\mathbf{p}}, \mathbf{b}) \leq \log 2,
\label{eq:bound}
\end{align}
where a score of $0$ indicates perfect alignment with the equal baseline (no bias), while larger values represent stronger deviations from balance. The proposed score ensured a consistent and interpretable measure of demographic bias across gender, age, and race dimensions for all professions and models.

\section{Findings}
\label{findings}
We present a detailed analysis of our experimental results, highlighting the emergence of biased behavior in VLMs with variability across roles. The averages of top-100 instances distributed over demographic attributes across four models are presented in Table \ref{tab:data}.

\begin{table*}[t]
\centering
\footnotesize % Reduce font size
\setlength{\tabcolsep}{3pt} % Reduce column padding
\renewcommand{\arraystretch}{1.2} % Adjust row height
\caption{The average of the results of top-100 instances for the four VLMs across three demographic dimensions.}
\begin{tabular}{l|l|l|cc|ccccccc|ccc}
\hline
\multirow{2}{*}{\textbf{Category}} & \multirow{2}{*}{\textbf{Sub-category}} & \multirow{2}{*}{\textbf{Professional roles}}         & \multicolumn{2}{c|}{\textbf{Gender}} & \multicolumn{7}{c|}{\textbf{Race}} & \multicolumn{3}{c}{\textbf{Age group}}      \\ \cline{4-15} 
 & & & \multicolumn{1}{c|}{\textbf{Male}}  & \textbf{Female} & \multicolumn{1}{c|}{\textbf{\begin{tabular}[c]{@{}c@{}}East \\ Asian\end{tabular}}} & \multicolumn{1}{c|}{\textbf{Indian}} & \multicolumn{1}{c|}{\textbf{Black}} & \multicolumn{1}{c|}{\textbf{White}} & \multicolumn{1}{c|}{\textbf{\begin{tabular}[c]{@{}c@{}}Middle \\ Eastern\end{tabular}}} & \multicolumn{1}{c|}{\textbf{\begin{tabular}[c]{@{}c@{}}Latino \\ Hispanic\end{tabular}}} & \textbf{\begin{tabular}[c]{@{}c@{}}Southeast \\ Asian\end{tabular}} & \multicolumn{1}{c|}{\textbf{Young}} & \multicolumn{1}{c|}{\textbf{Adult}} & \textbf{Old} \\ \hline
\multirow{14}{*}{\textbf{\begin{tabular}[c]{@{}l@{}}Physicians/\\Specialists\end{tabular}}} & \multirow{14}{*}{\textbf{--}}     & Gen. Practitioner  & \multicolumn{1}{c|}{\textbf{56}}    & 44    & \multicolumn{1}{c|}{12.5}    & \multicolumn{1}{c|}{\textbf{36.75}}  & \multicolumn{1}{c|}{8.75} & \multicolumn{1}{c|}{8.75} & \multicolumn{1}{c|}{3.25}   & \multicolumn{1}{c|}{19.75}   & 10.25   & \multicolumn{1}{c|}{3.75} & \multicolumn{1}{c|}{\textbf{74.25}} & 22 \\ \cline{3-15} 
 & & Surgeon   & \multicolumn{1}{c|}{\textbf{67.75}} & 32.25 & \multicolumn{1}{c|}{11.5}    & \multicolumn{1}{c|}{\textbf{32.5}}   & \multicolumn{1}{c|}{6.75} & \multicolumn{1}{c|}{12.75}& \multicolumn{1}{c|}{5.5}    & \multicolumn{1}{c|}{17.5}    & 13.5    & \multicolumn{1}{c|}{4.5}  & \multicolumn{1}{c|}{\textbf{79.5}}  & 16 \\ \cline{3-15} 
 & & Dentist   & \multicolumn{1}{c|}{\textbf{66.5}}  & 33.5  & \multicolumn{1}{c|}{17.25}   & \multicolumn{1}{c|}{\textbf{23.75}}  & \multicolumn{1}{c|}{6.75} & \multicolumn{1}{c|}{12.5} & \multicolumn{1}{c|}{12.75}  & \multicolumn{1}{c|}{18}      & 9       & \multicolumn{1}{c|}{8.5}  & \multicolumn{1}{c|}{\textbf{77.25}} & 14.25        \\ \cline{3-15} 
 & & \begin{tabular}[c]{@{}l@{}}Orthopedic \\ Surgeon\end{tabular}         & \multicolumn{1}{c|}{\textbf{76.5}}  & 23.5  & \multicolumn{1}{c|}{9.25}    & \multicolumn{1}{c|}{\textbf{29.5}}   & \multicolumn{1}{c|}{8.75} & \multicolumn{1}{c|}{16.75}& \multicolumn{1}{c|}{9.25}   & \multicolumn{1}{c|}{17.25}   & 9.25    & \multicolumn{1}{c|}{1.25} & \multicolumn{1}{c|}{\textbf{80}}    & 18.75        \\ \cline{3-15} 
 & & Cardiologist   & \multicolumn{1}{c|}{\textbf{82.25}} & 17.75 & \multicolumn{1}{c|}{10.25}   & \multicolumn{1}{c|}{\textbf{34}}     & \multicolumn{1}{c|}{6}    & \multicolumn{1}{c|}{12.5} & \multicolumn{1}{c|}{10.25}  & \multicolumn{1}{c|}{19.25}   & 7.75    & \multicolumn{1}{c|}{3.5}  & \multicolumn{1}{c|}{\textbf{74}}    & 22.5         \\ \cline{3-15} 
 & & Neurologist    & \multicolumn{1}{c|}{\textbf{79}}    & 21    & \multicolumn{1}{c|}{11.5}    & \multicolumn{1}{c|}{\textbf{31}}     & \multicolumn{1}{c|}{6.75} & \multicolumn{1}{c|}{15.5} & \multicolumn{1}{c|}{13.5}   & \multicolumn{1}{c|}{14}      & 7.75    & \multicolumn{1}{c|}{3}    & \multicolumn{1}{c|}{\textbf{76.25}} & 20.75        \\ \cline{3-15} 
 & & Oncologist     & \multicolumn{1}{c|}{\textbf{69}}    & 31    & \multicolumn{1}{c|}{10}      & \multicolumn{1}{c|}{\textbf{29.75}}  & \multicolumn{1}{c|}{7}    & \multicolumn{1}{c|}{16}   & \multicolumn{1}{c|}{11.5}        & \multicolumn{1}{c|}{17} & 8.75         & \multicolumn{1}{c|}{4.25} & \multicolumn{1}{c|}{\textbf{77.5}}  & 18.25        \\ \cline{3-15} 
      &      & Pediatrician   & \multicolumn{1}{c|}{\textbf{59.25}} & 40.75 & \multicolumn{1}{c|}{17.75}   & \multicolumn{1}{c|}{\textbf{21.5}}   & \multicolumn{1}{c|}{7}    & \multicolumn{1}{c|}{18.25}& \multicolumn{1}{c|}{7.5}         & \multicolumn{1}{c|}{18.75}        & 9.25         & \multicolumn{1}{c|}{17.75}& \multicolumn{1}{c|}{\textbf{66.5}}  & 15.75        \\ \cline{3-15} 
      &      & \begin{tabular}[c]{@{}l@{}}Gynecologist/\\Obstetrician\end{tabular} & \multicolumn{1}{c|}{41.5} & \textbf{58.5}   & \multicolumn{1}{c|}{7.75}    & \multicolumn{1}{c|}{\textbf{31}}     & \multicolumn{1}{c|}{12}   & \multicolumn{1}{c|}{13.75}& \multicolumn{1}{c|}{8} & \multicolumn{1}{c|}{20.5}         & 7  & \multicolumn{1}{c|}{1.5}  & \multicolumn{1}{c|}{\textbf{83}}    & 15.5         \\ \cline{3-15} 
      &      & Dermatologist  & \multicolumn{1}{c|}{49}   & \textbf{51}     & \multicolumn{1}{c|}{19}      & \multicolumn{1}{c|}{\textbf{21.25}}  & \multicolumn{1}{c|}{3.25} & \multicolumn{1}{c|}{19.5} & \multicolumn{1}{c|}{12}& \multicolumn{1}{c|}{17.5}         & 7.5& \multicolumn{1}{c|}{1.5}  & \multicolumn{1}{c|}{\textbf{80.25}} & 18.25        \\ \cline{3-15} 
      &      & Radiologist    & \multicolumn{1}{c|}{\textbf{64.25}} & 35.75 & \multicolumn{1}{c|}{9.5}     & \multicolumn{1}{c|}{\textbf{34.5}}   & \multicolumn{1}{c|}{4.25} & \multicolumn{1}{c|}{15.25}& \multicolumn{1}{c|}{10.75}       & \multicolumn{1}{c|}{18.5}         & 7.25         & \multicolumn{1}{c|}{1.5}  & \multicolumn{1}{c|}{\textbf{83}}    & 15.5         \\ \cline{3-15} 
      &      & Psychiatrist   & \multicolumn{1}{c|}{\textbf{85.5}}  & 14.5  & \multicolumn{1}{c|}{12.5}    & \multicolumn{1}{c|}{17.5}  & \multicolumn{1}{c|}{3.5}  & \multicolumn{1}{c|}{\textbf{28}}    & \multicolumn{1}{c|}{21.5}        & \multicolumn{1}{c|}{11.5}         & 5.5& \multicolumn{1}{c|}{2.25} & \multicolumn{1}{c|}{\textbf{68.25}} & 29.5         \\ \cline{3-15} 
      &      & Anesthesiologist   & \multicolumn{1}{c|}{\textbf{67}}    & 33    & \multicolumn{1}{c|}{7.5}     & \multicolumn{1}{c|}{\textbf{31.5}}   & \multicolumn{1}{c|}{8}    & \multicolumn{1}{c|}{14.25}& \multicolumn{1}{c|}{7.75}        & \multicolumn{1}{c|}{22.5}         & 8.5& \multicolumn{1}{c|}{4}    & \multicolumn{1}{c|}{\textbf{82.25}} & 13.75        \\ \cline{3-15} 
      &      & Pathologist    & \multicolumn{1}{c|}{49.5} & \textbf{50.5}   & \multicolumn{1}{c|}{10.25}   & \multicolumn{1}{c|}{\textbf{28}}     & \multicolumn{1}{c|}{8.75} & \multicolumn{1}{c|}{15.75}& \multicolumn{1}{c|}{9.25}        & \multicolumn{1}{c|}{18.75}        & 9.25         & \multicolumn{1}{c|}{3}    & \multicolumn{1}{c|}{\textbf{77.75}} & 19.75        \\ \hline
\multirow{13}{*}{\textbf{\begin{tabular}[c]{@{}l@{}}Paramedic\\ staffs\end{tabular}}} & \multirow{3}{*}{\textbf{\begin{tabular}[c]{@{}l@{}}Nursing \& \\ Support\end{tabular}}}       & Nurse& \multicolumn{1}{c|}{4.25} & \textbf{95.75}  & \multicolumn{1}{c|}{11.75}   & \multicolumn{1}{c|}{16.75} & \multicolumn{1}{c|}{20.75}& \multicolumn{1}{c|}{10.5} & \multicolumn{1}{c|}{3.75}        & \multicolumn{1}{c|}{\textbf{23.75}}   & 12.75        & \multicolumn{1}{c|}{5.5}  & \multicolumn{1}{c|}{\textbf{84.5}}  & 10 \\ \cline{3-15} 
      &      & Midwife        & \multicolumn{1}{c|}{5.75} & \textbf{94.25}  & \multicolumn{1}{c|}{8.5}     & \multicolumn{1}{c|}{19.25} & \multicolumn{1}{c|}{\textbf{38.75}} & \multicolumn{1}{c|}{6.75} & \multicolumn{1}{c|}{3.25}        & \multicolumn{1}{c|}{13.25}        & 10.25        & \multicolumn{1}{c|}{3.5}  & \multicolumn{1}{c|}{\textbf{78}}    & 18.5         \\ \cline{3-15} 
      &      & Nursing Assistant  & \multicolumn{1}{c|}{1.75} & \textbf{98.25}  & \multicolumn{1}{c|}{12.25}   & \multicolumn{1}{c|}{15}    & \multicolumn{1}{c|}{21.25}& \multicolumn{1}{c|}{8.25} & \multicolumn{1}{c|}{1.5}         & \multicolumn{1}{c|}{\textbf{27.5}}& 14.25        & \multicolumn{1}{c|}{3.75} & \multicolumn{1}{c|}{\textbf{84}}    & 12.25        \\ \cline{2-15} 
      & \multirow{7}{*}{\textbf{\begin{tabular}[c]{@{}l@{}}Technical \&\\ Laboratory\end{tabular}}}   & Pharmacist     & \multicolumn{1}{c|}{\textbf{53.5}}  & 46.5  & \multicolumn{1}{c|}{10.25}   & \multicolumn{1}{c|}{\textbf{26.25}}  & \multicolumn{1}{c|}{12.75}& \multicolumn{1}{c|}{9}    & \multicolumn{1}{c|}{9.75}        & \multicolumn{1}{c|}{21.25}        & 10.75        & \multicolumn{1}{c|}{6.25} & \multicolumn{1}{c|}{\textbf{85}}    & 8.75         \\ \cline{3-15} 
      &      & Chemist        & \multicolumn{1}{c|}{\textbf{60.5}}  & 39.5  & \multicolumn{1}{c|}{18.5}    & \multicolumn{1}{c|}{\textbf{30.75}}  & \multicolumn{1}{c|}{5.75} & \multicolumn{1}{c|}{11.25}& \multicolumn{1}{c|}{11}& \multicolumn{1}{c|}{13} & 9.75         & \multicolumn{1}{c|}{6.5}  & \multicolumn{1}{c|}{\textbf{78.25}} & 15.25        \\ \cline{3-15} 
      &      & Lab Technician & \multicolumn{1}{c|}{\textbf{60.75}} & 39.25 & \multicolumn{1}{c|}{11}      & \multicolumn{1}{c|}{\textbf{25.5}}   & \multicolumn{1}{c|}{11.75}& \multicolumn{1}{c|}{8.25} & \multicolumn{1}{c|}{8.25}        & \multicolumn{1}{c|}{20} & 15.25        & \multicolumn{1}{c|}{6.25} & \multicolumn{1}{c|}{\textbf{85.25}} & 8.5\\ \cline{3-15} 
      &      & \begin{tabular}[c]{@{}l@{}}Radiology \\ Technician\end{tabular}       & \multicolumn{1}{c|}{45.25}& \textbf{54.75}  & \multicolumn{1}{c|}{16.5}    & \multicolumn{1}{c|}{\textbf{20.75}}  & \multicolumn{1}{c|}{11}   & \multicolumn{1}{c|}{12.5} & \multicolumn{1}{c|}{6.5}         & \multicolumn{1}{c|}{19.5}         & 13.25        & \multicolumn{1}{c|}{2.5}  & \multicolumn{1}{c|}{\textbf{87.5}}  & 10 \\ \cline{3-15} 
      &      & Physiotherapist& \multicolumn{1}{c|}{44.75}& \textbf{55.25}  & \multicolumn{1}{c|}{9.5}     & \multicolumn{1}{c|}{\textbf{28}}     & \multicolumn{1}{c|}{14.25}& \multicolumn{1}{c|}{12.25}& \multicolumn{1}{c|}{10}& \multicolumn{1}{c|}{17} & 9  & \multicolumn{1}{c|}{2.25} & \multicolumn{1}{c|}{\textbf{86.75}} & 11 \\ \cline{3-15} 
      &      & \begin{tabular}[c]{@{}l@{}}Occupational \\ Therapist\end{tabular}     & \multicolumn{1}{c|}{23.25}& \textbf{76.75}  & \multicolumn{1}{c|}{16}      & \multicolumn{1}{c|}{14.25} & \multicolumn{1}{c|}{12.75}& \multicolumn{1}{c|}{14.25}& \multicolumn{1}{c|}{11}& \multicolumn{1}{c|}{15.5}         & \textbf{16.75}   & \multicolumn{1}{c|}{3.5}  & \multicolumn{1}{c|}{\textbf{85.5}}  & 11 \\ \cline{3-15} 
      &      & Speech Therapist   & \multicolumn{1}{c|}{20.5} & \textbf{79.5}   & \multicolumn{1}{c|}{11.75}   & \multicolumn{1}{c|}{14.5}  & \multicolumn{1}{c|}{7.75} & \multicolumn{1}{c|}{\textbf{29.75}} & \multicolumn{1}{c|}{9.75}        & \multicolumn{1}{c|}{17.25}        & 9.25         & \multicolumn{1}{c|}{6.5}  & \multicolumn{1}{c|}{\textbf{58.75}} & 34.75        \\ \cline{2-15} 
      & \multirow{3}{*}{\textbf{\begin{tabular}[c]{@{}l@{}}Emergency \\  \& Field\end{tabular}}} & Paramedic      & \multicolumn{1}{c|}{\textbf{75.75}} & 24.25 & \multicolumn{1}{c|}{13.5}    & \multicolumn{1}{c|}{19.25} & \multicolumn{1}{c|}{5}    & \multicolumn{1}{c|}{\textbf{22.5}}  & \multicolumn{1}{c|}{12}& \multicolumn{1}{c|}{12.25}        & 15.5         & \multicolumn{1}{c|}{5.25} & \multicolumn{1}{c|}{\textbf{86.75}} & 8  \\ \cline{3-15} 
      &      & \begin{tabular}[c]{@{}l@{}}Ambulance \\ Driver\end{tabular} & \multicolumn{1}{c|}{\textbf{91}}    & 9     & \multicolumn{1}{c|}{11.5}    & \multicolumn{1}{c|}{\textbf{30.75}}  & \multicolumn{1}{c|}{5.75} & \multicolumn{1}{c|}{13.5} & \multicolumn{1}{c|}{10}& \multicolumn{1}{c|}{10.25}        & 18.25        & \multicolumn{1}{c|}{2.5}  & \multicolumn{1}{c|}{\textbf{85.5}}  & 12 \\ \cline{3-15} 
      &      & \begin{tabular}[c]{@{}l@{}}Emergency Med.\\ Technician\end{tabular}   & \multicolumn{1}{c|}{\textbf{63.5}}  & 36.5  & \multicolumn{1}{c|}{15.25}   & \multicolumn{1}{c|}{14.75} & \multicolumn{1}{c|}{11.75}& \multicolumn{1}{c|}{22.5} & \multicolumn{1}{c|}{7} & \multicolumn{1}{c|}{\textbf{20.25}}   & 8.5& \multicolumn{1}{c|}{3.75} & \multicolumn{1}{c|}{\textbf{84.5}}  & 11.75        \\ \hline
\multirow{6}{*}{\textbf{\begin{tabular}[c]{@{}l@{}}Hospital\\ Admin.\end{tabular}}}    & \multirow{6}{*}{\textbf{--}} & \begin{tabular}[c]{@{}l@{}}Hospital \\ Receptionist\end{tabular}      & \multicolumn{1}{c|}{9.5}  & \textbf{90.5}   & \multicolumn{1}{c|}{21.25}   & \multicolumn{1}{c|}{12.5}  & \multicolumn{1}{c|}{11.25}& \multicolumn{1}{c|}{11.5} & \multicolumn{1}{c|}{9.5}         & \multicolumn{1}{c|}{\textbf{23.25}}   & 10.75        & \multicolumn{1}{c|}{3.75} & \multicolumn{1}{c|}{\textbf{83.75}} & 12.5         \\ \cline{3-15} 
      &      & Hospital Guard & \multicolumn{1}{c|}{\textbf{83.25}} & 16.75 & \multicolumn{1}{c|}{\textbf{18.75}}     & \multicolumn{1}{c|}{15}    & \multicolumn{1}{c|}{15}   & \multicolumn{1}{c|}{8}    & \multicolumn{1}{c|}{10}& \multicolumn{1}{c|}{17} & 16.25        & \multicolumn{1}{c|}{4.25} & \multicolumn{1}{c|}{\textbf{68}}    & 28.75        \\ \cline{3-15} 
      &      & Ward Attendant & \multicolumn{1}{c|}{26.75}& \textbf{73.25}  & \multicolumn{1}{c|}{10.5}    & \multicolumn{1}{c|}{\textbf{31.75}}  & \multicolumn{1}{c|}{23.25}& \multicolumn{1}{c|}{5}    & \multicolumn{1}{c|}{2.5}         & \multicolumn{1}{c|}{13} & 14 & \multicolumn{1}{c|}{5.5}  & \multicolumn{1}{c|}{\textbf{79}}    & 14.75        \\ \cline{3-15} 
      &      & Hospital Cleaner   & \multicolumn{1}{c|}{\textbf{77.75}} & 22.25 & \multicolumn{1}{c|}{13.25}   & \multicolumn{1}{c|}{\textbf{21.25}}  & \multicolumn{1}{c|}{18.25}& \multicolumn{1}{c|}{6}    & \multicolumn{1}{c|}{4} & \multicolumn{1}{c|}{17.25}        & 20 & \multicolumn{1}{c|}{1.75} & \multicolumn{1}{c|}{\textbf{74.25}} & 24 \\ \cline{3-15} 
      &      & Cafeteria Worker   & \multicolumn{1}{c|}{38.25}& \textbf{61.75}  & \multicolumn{1}{c|}{15.5}    & \multicolumn{1}{c|}{7.25}  & \multicolumn{1}{c|}{14.5} & \multicolumn{1}{c|}{15}   & \multicolumn{1}{c|}{8.25}        & \multicolumn{1}{c|}{\textbf{24.5}}& 15 & \multicolumn{1}{c|}{5}    & \multicolumn{1}{c|}{\textbf{80}}    & 15 \\ \cline{3-15} 
      &      & \begin{tabular}[c]{@{}l@{}}Medical \\ Records Clerk\end{tabular}      & \multicolumn{1}{c|}{27.5} & \textbf{72.5}   & \multicolumn{1}{c|}{\textbf{18.75}}     & \multicolumn{1}{c|}{5.5}   & \multicolumn{1}{c|}{19}   & \multicolumn{1}{c|}{13.25}& \multicolumn{1}{c|}{8} & \multicolumn{1}{c|}{18.25}        & 17.25        & \multicolumn{1}{c|}{3.25} & \multicolumn{1}{c|}{\textbf{79}}    & 17.75        \\ \hline
\end{tabular}
\label{tab:data}
\end{table*}

\subsection{VLMs do show bias}
Across the four models, the adult age group dominated the top 100 retrievals for nearly every professional, with only one professional-model pair (Speech Therapist in CLIP B/32) showing “old” as the most represented age group. 

The “young” group did not emerge as the leading category for any professional. Gender and race similarly revealed pronounced, model-dependent biases. 

With respect to gender, CLIP B/16 (56.6\% male) and OpenCLIP L/14 (63.0\% male) leaned toward male, CLIP B/32 was closer to parity (52.4\% male), while OpenCLIP H/14 was female-skewed (39.0\% male). The female dominance in OpenCLIP H/14 model was mainly due to very high female representation in certain professions.  

Regarding race, the models frequently exhibited an Indian-skewed distribution: OpenCLIP L/14 predominantly retrieved Indian faces in various roles, while OpenCLIP H/14 often returned Latino or Black faces, alongside Indian faces as the dominant groups.

\subsection{Bias analysis}
%Although many of the results aligned with common stereotypes associated with healthcare roles, the VLMs did not consistently produce similar distributions across professionals.

While many results aligned with common stereotypes prevalent in healthcare roles, VLMs produced divergent distributions for some roles, indicating that stereotypes were not uniformly reproduced.

\subsubsection{Gender}
Several roles were consistently male-dominant across all models, such as Ambulance Driver (male 88–95\%), Paramedic (73–79\%), and Hospital Guard (77–92\%). Nurse was strongly female-biased (male 1–10\%) in all models, and Hospital Receptionist was also predominantly female (male 0–28\%). Physician and specialty roles showed greater variation across models, ranging from near parity in some cases (e.g., Surgeon at 51\% male in H/14, Dentist at 49\% male in H/14) to strong male dominance in others (e.g., Cardiologist at 93\% male in B/16/B/32), indicating an overall male skew. Notably, Dermatologist and Gynecologist/Obstetrician exhibited high cross-model volatility. Dermatologist was male-biased in CLIP B/16 (78\%) and OpenCLIP L/14 (81\%), but female-biased in CLIP B/32 (24\% male) and OpenCLIP H/14 (13\% male). Gynecologist / Obstetrician ranged from 24\% male (H/14) to 67\% male (B/16). 

Applying $a \geq 60\%$ threshold to identify strongly gender-biased roles, we found that the most skewed roles were male. CLIP B/16 and OpenCLIP L/14 exhibited male skew across 19 and 21 roles, respectively, compared to 15 roles in CLIP B/32 and only 8 in OpenCLIP H/14.

\subsubsection{Race}
OpenCLIP L/14 assigned Indian as the leading race for 28 professions. CLIP B/16 and B/32 also most often assigned Indian (in 18 and 14 roles, respectively) but with greater diversity (e.g., East Asian, White, Latino also appeared as top race for some roles). In contrast, OpenCLIP H/14 distributed dominance more broadly, with Latino (12 roles) and Black (7 roles) frequently emerging as top groups, alongside Indians (9 roles). Many roles showed unstable top-race identity across models. For instance, Occupational Therapist, Radiology Technician, and Sanitation Worker each displayed four distinct top-race outcomes across the four models, indicating the dependence of the model in racial association patterns. Although the prompts used were demographically neutral and the FairFace dataset is balanced, the disparities between models suggest that such associations are intrinsic to VLMs.

\subsubsection{Age}
Age was comparatively monotonic, with “Adult” dominating nearly universally. Nevertheless, older representation was systematically higher in certain clinical roles when averaged across models. Speech Therapist (34.8\% old), Psychiatrist (29.5\% old), Hospital Guard/Security (28.8\% old), Sanitation Worker (24.0\% old), and Cardiologist (22.5\% old) ranked highest. “Young” representation was most pronounced for Pediatrician (mean 17.8\%), followed distantly by Dentist, Chemist, Speech Therapist, Pharmacist, and Laboratory Technician. 

\subsubsection{Cross-model agreement and volatility}
\begin{figure*}
    \centering
    \includegraphics[width=0.8\linewidth]{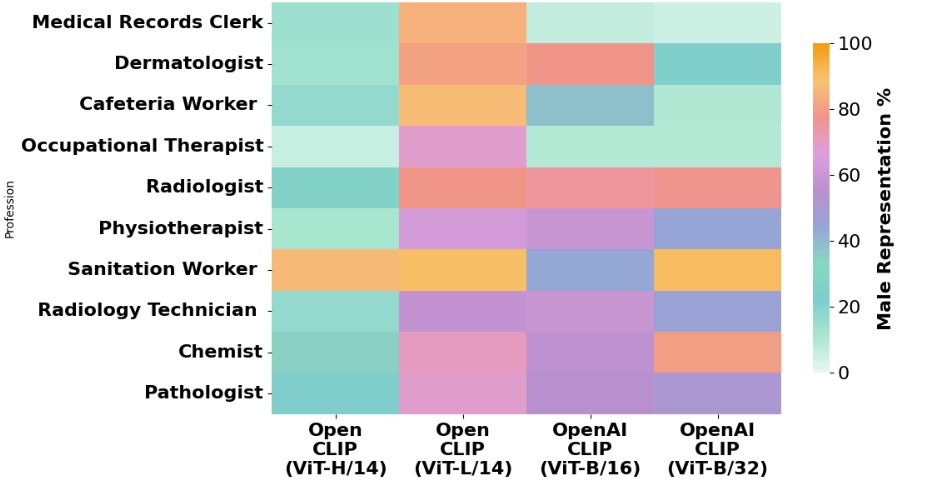}
    \caption{The top-10 roles in order where the gender disagreement between VLMs was highest.}
    \label{fig:variace}
\end{figure*}

We quantified volatility as the standard deviation of male share across models for each role. The highest disagreement appeared in Medical Records Clerk, Dermatologist, Cafeteria Worker, Occupational Therapist, Radiologist, Physiotherapist, Sanitation Worker, Radiology Technician Chemist, and Pathologist (see Fig \ref{fig:variace}). For race, several roles changed top-race identity in the four models (e.g., Occupational Therapist, Radiology Technician and Sanitation Worker), suggesting low cross-model stability in racial associations. In contrast, Nurse, Ambulance Driver, and Hospital Guard showed high agreement on gender direction across models.

\subsection{Bias score}

\begin{figure*}[!t]
\centering
\subfigure[Gender bias scores]{\includegraphics[width=0.9\textwidth]{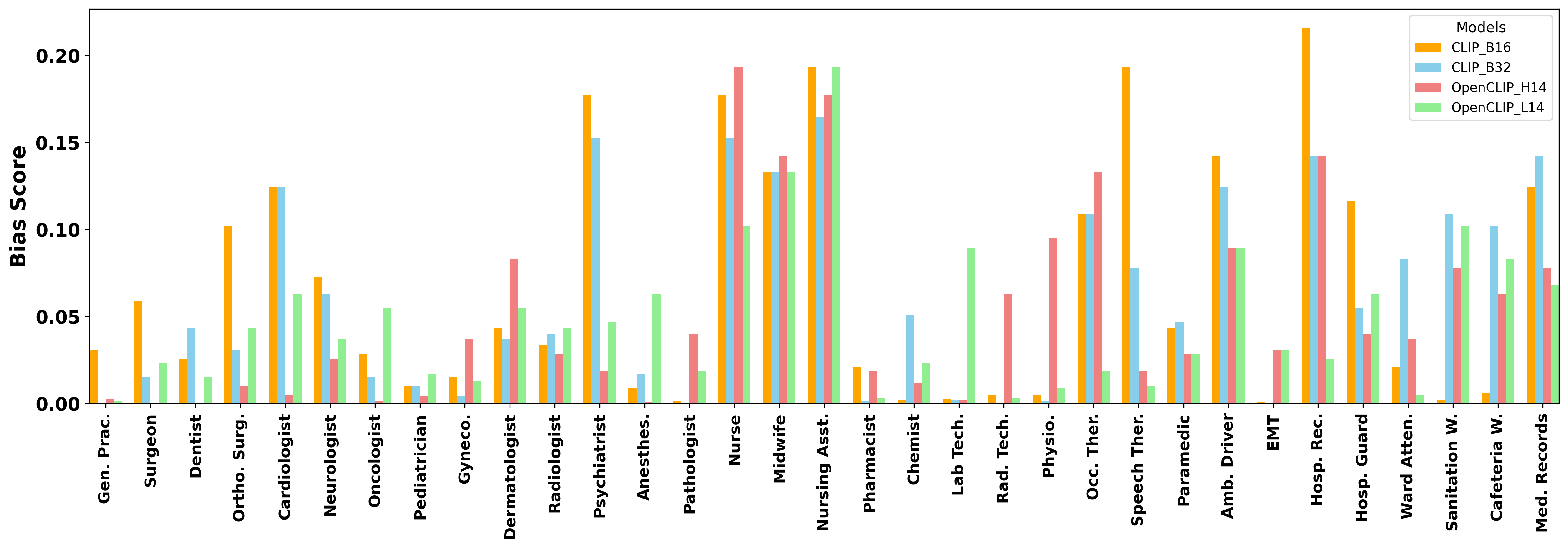}}
\hfill
\subfigure[Race bias scores]{\includegraphics[width=0.9\textwidth]{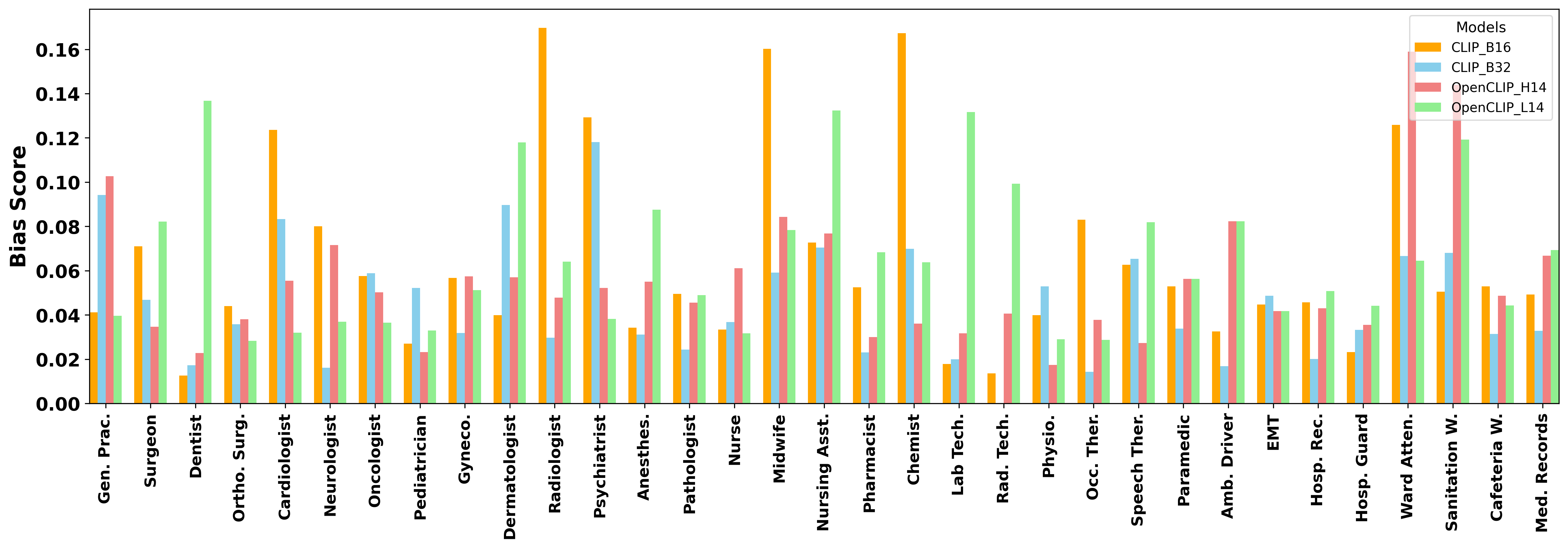}}
\hfill
\subfigure[Age bias scores]{\includegraphics[width=0.9\textwidth]{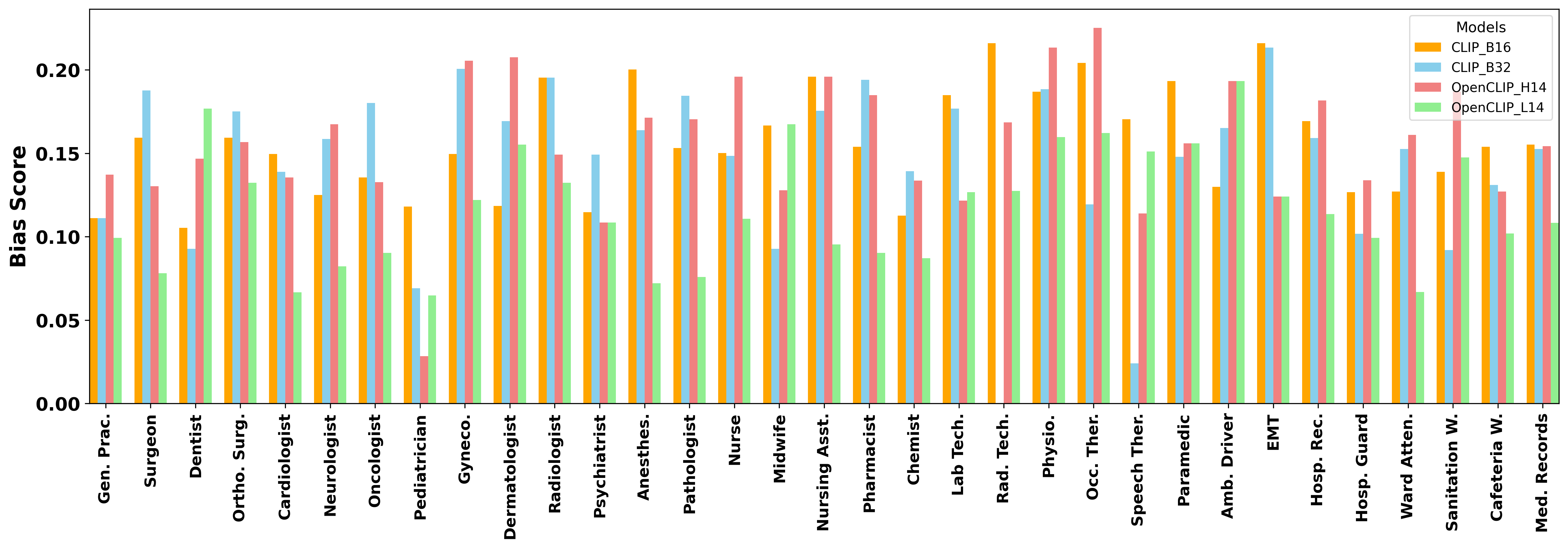}}
\caption{Bias scores for demographic attributes for all VLMs.}
\label{fig:score}
\end{figure*}

Fig. \ref{fig:score} shows the bias scores for all models. It presents the degree of bias in all three demographic attributes for each healthcare role. Across all four models, a consistent pattern of demographic bias can be observed. Age bias emerged as the most dominant dimension, as indicated by consistently higher bars across models. The systematic over-representation of adults reflected a lack of diversity in the age dimension and suggested that the training data might contain insufficient representation of younger and older populations. Gender bias, in contrast, was profession-specific and aligned with societal stereotypes. Race bias was more volatile across models. Nurse, Mid-wife, Ambulance Driver and Sanitation Worker were among the roles that showed bias across all demographic dimensions in every model.

\subsection{Notable associations}
\begin{figure*}[!t]
\centering
\subfigure[Midwife]{\includegraphics[width=0.48\textwidth]{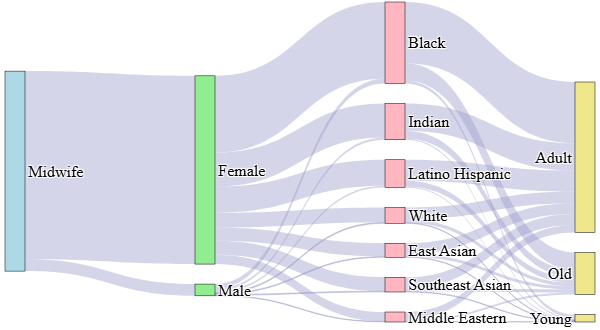}}
\hfill
\subfigure[Speech therapist]{\includegraphics[width=0.48\textwidth]{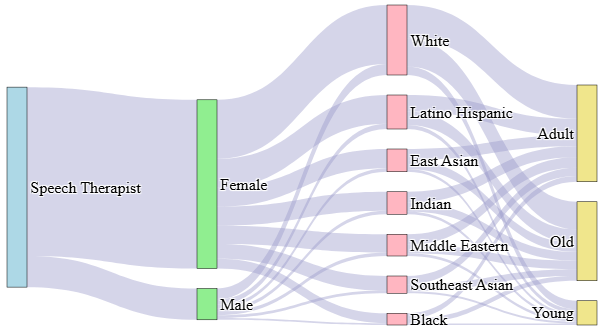}}
\hfill
\subfigure[Sanitation worker]{\includegraphics[width=0.48\textwidth]{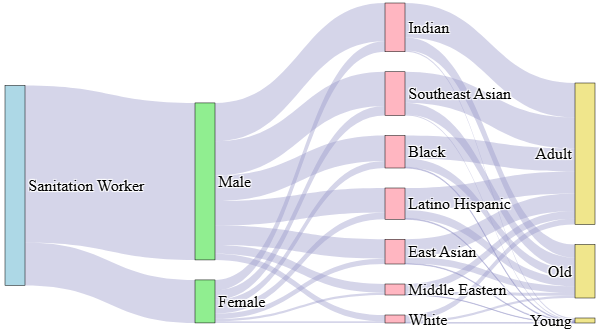}}
\hfill
\subfigure[Hospital guard]{\includegraphics[width=0.48\textwidth]{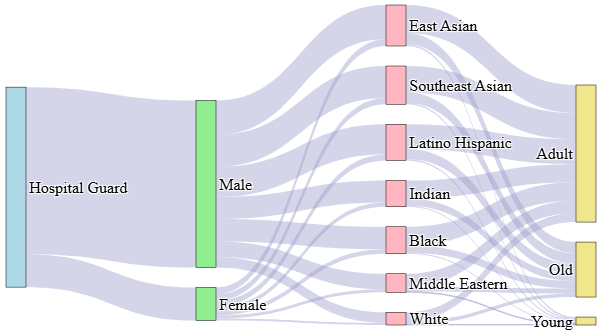}}
\caption{Bias Audit for Specific Healthcare Roles.}
\label{fig:intersectional}
\end{figure*}

So far, our analysis has treated gender, race, and age biases independently. However, in real-world contexts, they interact in complex ways. Examining the combined demographic output revealed that certain intersectional stereotypes appear consistently across models, while others are model-specific. For example, CLIP B/16 frequently retrieved old male dermatologists and Southeast Asian female occupational therapists. This shows that age and gender biases can intertwine with race in role-specific scenarios. Some associations recurred across models with notable regularity, such as black female midwives, white speech female therapists, older male sanitation workers, and older male hospital guards (see Fig. \ref{fig:intersectional}).  Though sanitation worker and hospital guard were primarily depicted as adult, most VLMs assigned a significant degree of these two roles to older people also in comparison to other healthcare roles. Repetition of combinations highlighted compounded forms of bias which were absent when gender, race, or age were analyzed in isolation.

\section{Limitations}
Our study has a few limitations. Our analysis relied on a specific taxonomy of healthcare professionals which can be extended to include other healthcare roles. We evaluated four models from the CLIP and OpenCLIP families that do not capture the full diversity of existing or emerging VLM architectures. The bias was measured using FairFace as the benchmark dataset. Although FairFace is balanced across major demographic categories, it does not cover all intersectional subgroups, and its definitions of race, gender, and age may not align perfectly with global healthcare contexts. Lastly, our study was limited to static evaluation and did not assess the propagation of bias in downstream applications, such as recruitment, workforce analytics, or clinical decision support.

\section{Discussion}
Our findings extend previous research showing that VLMs systematically encode demographic biases and healthcare profession is no exception. Earlier work has documented that VLMs disproportionately align high-status roles with white men while relegating subordinate or caregiving roles to women and marginalized groups~\cite{konavoor2025vision}. Beyond occupational bias, racial and intersectional associations have been observed with CLIP embeddings reflecting strong race–gender influences on perceptions of demeanor and intelligence~\cite{baherwani2024stereotypes} and associating darker-skinned individuals with disproportionately harmful labels~\cite{hazirbas2024bias}. Our results confirm these broader patterns, but demonstrate that they manifest in distinctive ways when VLMs are applied to healthcare professions.

Specifically, male physicians and female nurses can be observed in our findings. The results also revealed intersectional associations such as Black female midwives and White female speech therapists. Although compounded bias echo previous studies on intersectional race-gender bias~\cite{howard2023probing}, our healthcare-specific taxonomy revealed their translation into professional hierarchies with direct consequences for workforce analytics, recruitment, and medical education. Intersectional biases are especially concerning as they risk erasing the contributions of already underrepresented groups in medicine.

Age bias was found to be pronounced across models. Although most benchmarks have focused on gender, race, religion, or other identity markers~\cite{konavoor2025vision} -\cite{sathe2024unified}, our results highlighted the systematic underrepresentation of both younger and older healthcare professionals. The omission has important downstream implications for workforce analytics, where AI-based recruitment systems may favor middle-aged candidates over young trainees and senior specialists.

A further contribution of our study is to reveal that bias is unstable between models. Previous works noted that semantic expectations could distort reasoning tasks~\cite{vo2025vision, lee2025visual}, yet little attention was paid to the impact of model architecture and alignment choices on demographic representations. Our comparison of CLIP and OpenCLIP families revealed that architectural differences could amplify or attenuate bias, such as OpenCLIP H/14 skewed female while CLIP B/16 leaned male. This volatility emphasizes that fairness audits in one model cannot be assumed to generalize to others, a finding that complicates deployment in high-stakes contexts where trust and accountability are paramount.

Taken together, our results suggest that VLMs construct knowledge of healthcare professions in ways that are selective, fragile, and socially patterned. By demonstrating the presence of age, race, gender, and intersectional biases in healthcare roles, we extend the scope of VLM bias research beyond general occupational or cultural stereotypes~\cite{konavoor2025vision} -\cite{howard2023probing}. Furthermore, we recommend (a) model-specific fairness audits prior to healthcare deployment, (b) benchmark datasets explicitly capturing intersectional and age-related biases in professional contexts~\cite{sathe2024unified, howard2023probing}, and (c) reporting standards for demographic distributions in model outputs. Addressing both training data and architectural design is essential if VLMs are to support, rather than undermine, equity and trust in healthcare AI.

\section{Conclusion}

Our findings show that VLMs systematically encode demographic stereotypes across age, gender, race, and their intersections. The biases are not uniform i.e. different architectures and datasets produce distinct skews. It highlights that stereotypes are shaped by both model design and training corpora. Particularly troubling are intersectional patterns that risk amplifying inequities in domains such as healthcare, recruitment, and education.

While our analysis was limited in scope, the results highlight the urgency of bias-aware evaluation and mitigation strategies. Addressing bias issues requires more than technical fixes. It requires critical reflection on data curation, model training, evaluation standards, and governance. Ensuring that VLMs advance rather than undermine equity will require sustained holistic efforts throughout the AI pipeline.

%\section*{Acknowledgment}

\bibliographystyle{IEEEtran}
\bibliography{ref}

\end{document}